\documentstyle[aps,twocolumn]{revtex}
\input{epsf}
\def\beq{\begin{equation}}
\def\eeq{\end{equation}}
\def\bea{\begin{eqnarray}}
\def\eea{\end{eqnarray}}
\def\ba{\begin{array}}
\def\ea{\end{array}}

\def\gev{1\,{\rm GeV}}
\def\nue{{\nu_{e}}}
\def\numu{{\nu_{\mu}}}
\def\nutau{\nu_{\tau}}

\begin{document}

%MADPH-98-1077 \\
%UH-511-916-98 \\
%VAN D-TH-98-17}}}
%\end{quote}
%\end{flushright}
%\vspace{.10in}

\title{Neutrino Decay as an Explanation of\\ Atmospheric Neutrino Observations}

\author{$^{1,4}$V.~Barger, $^2$J.~G.~Learned, $^2$S.~Pakvasa, and $^3$T.~J.~Weiler}
\address{$^1$Department of Physics, University of Wisconsin, Madison, WI 53706, USA, \\
$^2$Department of Physics and Astronomy, University of Hawaii, Honolulu, HI 96822, USA, \\
$^3$Department of Physics and Astronomy, Vanderbilt University, Nashville, TN 37235, USA \\
$^4$Fermi National Accelerator Laboratory, Batavia, IL  60510, USA}

\maketitle

\begin{abstract}
We show that the observed zenith angle dependence of the atmospheric neutrinos
can be accounted for by neutrino decay. Furthermore, it is possible to account
for all neutrino anomalies with just three flavors.
\end{abstract}
%\thispagestyle{empty}

%\newpage

\section{Neutrino Decay Phenomenology}

According to the analysis by the Super-Kamiokande (SK)
Collaboration\cite{super-k} of their atmospheric neutrino data, the
$L/E$ dependence of the $\nu_\mu$ data is well accounted for by
$\nu_\mu$ oscillation into $\nu_\tau$ (or a sterile neutrino
$\nu_{st}$) with a mixing angle $\sin^22\theta\agt0.8$ and a $\delta
m^2$ in the range $3\times10^{-4}$--$8\times10^{-3}\rm\,eV^2$. The
$\nu_\mu$ survival probability in vacuum is given by

\begin{equation}
P_{\mu\mu} = 1 - \sin^2 2\theta\,\sin^2(\delta m^2 L/4E)\,.
\label{Pmumu-osc}
\end{equation}

According to the data and in this simple two flavor mixing
hypothesis, the $\nu_e$ channel is unaffected and hence the survival
probability $P_{ee} = 1$. The interpretation of the atmospheric
neutrino data in terms of oscillations is an old and venerable
proposal\cite{old}.

It is obviously an important question to ask whether $\nu_\mu$
oscillation is the only possible explanation for the observed $L/E$
dependence. Several other interpretations have been
offered\cite{ma}. In this Letter we raise another possibility -
neutrino decay.

We begin by noting that decay implies a non-zero mass difference
between two neutrino states and thus, in general, mixing as
well. For definiteness, let us assume the existence of just three
light neutrinos, and label as $\nu_1$, $\nu_2$, and $\nu_3$ that
mass eigenstate with the largest admixture in the flavor state
$\nue$, $\numu$, and $\nutau$, respectively.  We further assume the
dominant component of $\numu$, i.e. $\nu_2$, to be the only unstable
state, with a rest-frame lifetime $\tau_0$.  There are strong limits
coming from the nonobservation of $\pi$ and $K$ decay to anomalous
final states containing $e^\pm$'s on the participation of $\nue$ in
non--SM vertices\cite{britton}.  Consequently, $\nue$ must nearly
decouple from the unstable $\nu_2$ and its decay partner $\nu_3$,
and we are led to $\nue\approx \nu_1$, and

\begin{equation}
\nu_\mu \approx \cos\theta\,\nu_2 + \sin\theta\,\nu_3 \,,
\label{numumix}
\end{equation}

with $m_2 > m_3$.  From Eq.\ (\ref{numumix}) with an unstable
$\nu_2$, the $\nu_\mu$ survival probability is

\begin{eqnarray}
P_{\mu\mu} &=& \sin^4\theta + \cos^4\theta \exp(-\alpha L/E)\nonumber\\
&& {}+ 2\sin^2\theta \cos^2\theta \exp(-\alpha L/2E) \cos(\delta m^2 L/2E)\,,
\label{eq:PmumuNoave}
\end{eqnarray}

where $\delta m^2 = m_2^2 - m_3^2$ and $\alpha = m_2/\tau_0$.

If, as we argue later, $\delta m^2 > 0.1\rm~eV^2$, then $\cos(\delta
m^2 L/2E)$ effectively averages to zero for atmospheric neutrinos
and $P_{\mu\mu}$ becomes

\begin{equation}
P_{\mu\mu} = \sin^4\theta + \cos^4\theta \exp(-\alpha L/E) \,.
\label{eq:Pmumu}
\end{equation}

\begin{figure}[htb]
\centerline{\epsfysize 3.5 truein 
\epsfbox{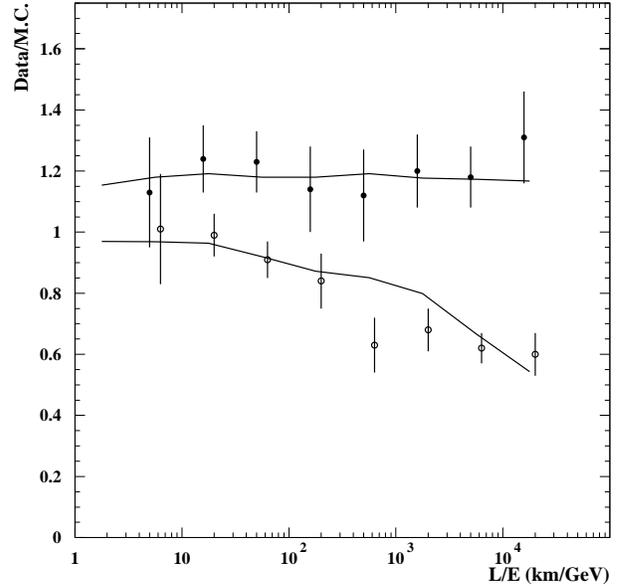}}
\caption{The Super-Kamiokande data/expectations as a function of
L/E, for electron events (upper) and muon events (lower). Our model
normalized to the electron flux total is shown by the lines,
indicating an acceptable fit for decaying muon neutrinos.}
\label{fig:loe} 
\end{figure}

We find that it is possible to choose $\theta$ and $\alpha$ to
provide a good fit to the Super-K $L/E$ distributions of $\nu_\mu$
events and $\numu/\nue$ event ratio.  The fit to the $L/E$
distribution of the $\numu$ events is shown in Figure \ref{fig:loe},
where smearing over L and E is included.  The best fit values of the
two parameters are $\cos^2\theta \sim 0.97$
(i.e. $\theta\sim10^\circ$) and $\alpha \sim \gev/D_E$, where $D_E =
12,800$ km is the diameter of the Earth.  This best--fit $\alpha$
value corresponds to a rest-frame $\nu_2$ lifetime of

\begin{equation}
\tau_0 =m_2/\alpha\sim {m_2\over(1{\rm\ eV})} \times 10^{-10}\rm\  s \ .
\label{RFlifetime}
\end{equation}

Such a lifetime and decay length are not in conflict with current
limits on non-radiative modes\cite{pdg98}.  Previous limits on
$\numu$ decay lengths from accelerators are only of the order of a
few km.

\section{Non--Radiative Neutrino--Decay Models}

The simplest possibility for fast invisible neutrino decay is to
choose $\nu_2$ and $\nu_3$ to be Dirac states, with a decay
interaction between the right--handed $SU(2)$--singlet states given
by\cite{acker}

\begin{equation}
{\cal L}_I = g_{23}\,\overline{\nu_{3R}^c}\,\nu_{2R}\,\chi + \rm h.c.\,,
\label{Dintn}
\end{equation}

where $\chi$ is a complex scalar field with lepton number $-2$,
$I_W=0$, and hypercharge zero.  $\chi$ is not a Nambu-Goldstone
boson.  The scalar field should be light compared to the neutrino
masses.

The quantum numbers of $\chi$ make it difficult to test its existence
in laboratory experiments.
Processes such as
$\mu\to e+ X$ are forbidden.
Furthermore, neutrino decay is not helicity--suppressed
but the interactions of its decay products are:
all new processes involving neutrinos
are suppressed by the chirality factor $(m_\nu/E_\nu)^2$.

A decay model for Majorana neutrinos can also be constructed.
The effective decay interaction is
\beq
{\cal L}_I = g_{23}\,\overline{\nu_{3L}^c}\,\nu_{2L}\,J + \rm h.c.\,,
\label{Mintn}
\end{equation}

where $J$ is a Majoron field\cite{gelmini}, which has to be
dominantly singlet with only a small triplet admixture, in order to
satisfy the constraints from the invisible decay width of $Z$. A
full model needs futher embellishment to generate the above coupling
at tree level and generate the desired neutrino masses and mixings.
The model described by Acker et al.\cite{MajModel} can be easily
adapted to our purpose here.

With the interaction of Eq.\ (\ref{Dintn}) or Eq.\ (\ref{Mintn}),
the rest-frame lifetime of $\nu_2$ is given by

\begin{equation}
\tau_0 = {16\pi\over g_{23}^2} {m_2^3 \over\delta m^2 (m_2+m_3)^2 }\,,
\end{equation}
\label{modeltime}

and hence

\beq
g_{23}^2\,\delta m^2 = 16\pi\alpha\,(1+m_3/m_2)^{-2}\,,
\eeq

leading to, for $0 < m_3/m_2 < 1$,

\begin{equation}
g_{23}^2 \,\delta m^2 \sim(2- 7)\times 10^{-4}\rm\, eV^2 \,.
\end{equation}

From studying $K$ decays, in particular the modes $K\to \mu +
neutrals$, a bound on the coupling $g_{23}$ can be derived which
is\cite{bkp}

\begin{equation}
g_{23}^2 < 2.4\times 10^{-4} \,,
\end{equation}

thus leading to a bound on $\delta m^2$ in the present model of

\begin{equation}
\delta m^2_{23} \agt 0.73\rm\ eV^2 \,.
\label{deltamsq23}
\end{equation}

This result justifies the above approximation of large  $\delta m^2$.

In both the Dirac neutrino and the Majorana neutrino decay models,
the small mass scale for the scalar fields seems to require
fine--tuning.  This is unavoidable, though unappealing.  The Dirac
and Majorana models are testably different in that in the
Dirac--neutrino model the decay is into essentially undetectable
final states, whereas the Majorana--neutrino model posits decay of
$\numu$'s and $\bar\numu$'s into $\bar\nutau$'s and $\nutau$'s
respectively, which are observable in principle.  Furthermore, with
the Majorana--neutrino model, lepton number is broken and
neutrinoless double--beta decay is allowed; with the Dirac--neutrino
model, lepton number is unbroken and neutrinoless double--beta decay
is not allowed.

If the decay of the $\nu_2$ is into a new(sterile) neutrino with
which it does not mix, then the $\delta m^2$ appearing in the
oscillation is not restricted by the constraint of Eq. (12) and can
be very small. Then the cosine term in Eq. (3) is essentially unity
and the survival probability is given by

\begin{eqnarray}
P_{\mu\mu} &=& [\sin^2\theta  + \cos^2\theta \exp(-\alpha L/2E)]^2.
\label{eq:Pmumu0}
\end{eqnarray}

In this case, even better fits to the Super-K data can be obtained.
However, we do not pursue this class of models further at this time.

\section{Inclusion of All Neutrino Anomalies}

It is easy to incorporate the solar and LSND neutrino anomalies into
the present discussion. The decay of $\nu_2$ with $\delta
m^2_{23}\agt 0.73 {\rm eV}^2$ explains the atmospheric data, but
also allows $\delta m^2_{23}$ to accommodate the LSND result.  We
are free to choose the remaining $\delta m^2$ to accommodate the
solar anomaly.  As mentioned earlier, $U_{e3}$ must be rather small;
and hence the solution to the solar $\nue$ depletion is necessarily
small--angle mixing enhanced by the MSW effect.  To this end, we set
$\delta m^2_{sun}\equiv\delta m_{31}^2 \sim 0.5\times
10^{-5}\rm\,eV^2$ to complement the $\delta m^2_{23}$ of Eq.\
(\ref{deltamsq23}).

\medskip\noindent
\underline{\it Atmospheric $\nu$'s}

With these $\delta m_{ij}^2$, the $\nu_\mu$ survival probability is
given by Eq.~(\ref{eq:Pmumu}) with $\cos^2\theta \equiv |U_{\mu
2}|^2\sim 0.97$ and $\sin^2\theta = 1-|U_{\mu 2}|^2$, and the
earlier two-flavor discussion and fit remain intact. The survival
probability for $\nu_e$'s is

\begin{eqnarray}
P_{ee} &=& \left(1-|U_{e2}|^2\right)^2 + |U_{e2}|^2 e^{-\alpha L/E}\nonumber 
\\ && {}+ 2|U_{e2}|^2(1-|U_{e2}|^2) \, \cos(\delta m_{32}^2 L/2E) e^{-\alpha L/2E}\,.
\end{eqnarray}

If $U_{e2} \ll 1$, then $P_{ee}\approx 1$, as observed.

\medskip\noindent
\underline{\it Solar $\nu$'s}

For solar $L/E$ values, the $\nu_2$'s have decayed away and the
$\nu_e$ survival probability is given by

\begin{equation}
P_{ee} = \left(1-|U_{e2}|^2\right)^2 - 4|U_{e1}|^2 |U_{e3}|^2 \sin^2
\left(\delta
m_{31}^2 L\over 4E\right) \,.
\end{equation}

By choosing $4|U_{e1}|^2 |U_{e3}|^2 \approx 5.5\times 10^{-3}$, with
$m_3 > m_1$, one can reproduce the small--angle MSW solution for
solar neutrinos\cite{bahcall}. Furthermore, the resulting value of
$|U_{e3}|^2\sim 1.4\times 10^{-3}$ easily satisfies the upper bound
on the $\nu_e$ coupling which is $g_e^2=g_{23}^2(|U_{e3}|^2 + |U_{e
2}|^2) < 3\times 10^{-5}$\cite{britton}.

\medskip\noindent
\underline{\it LSND}

At the $L/E$ value relevant to the LSND experiment, the
$\nu_\mu$--$\nu_e$ conversion probability is given by

\begin{equation}
P_{\mu e} = 4|U_{\mu 2}|^2 |U_{e2}|^2 \sin^2\left(\delta m_{32}^2 L\over
4E\right)\,.
\end{equation}

With $\delta m_{23}^2 \sim {\cal O}(1\rm\ eV)^2$ and
$|U_{\mu2}|^2\sim 1$, choosing $|U_{e2}|^2\sim 10^{-3}$ allows this
$P_{\mu e}$ to account for the LSND observations\cite{lsnd}.

\medskip\noindent
\underline{\bf Summary of Mass and Mixing}

To summarize, the three--neutrino mixing matrix with the approximate form

\begin{equation}
U = \left(\begin{array}{ccc}
0.999& 0.02& -0.04\\
\ -0.05& 0.985& 0.17 \\
\epsilon& \ -0.17& 0.985
\end{array}\right)
\label{U}
\end{equation}

and mass differences given by

\begin{equation}
\delta m_{21}^2 \approx \delta m_{23}^2 \agt 0.73\rm\ eV^2\,,
\hspace{1.0mm}{\rm and}\hspace{1.0mm}
\delta m_{13}^2 \sim 0.5\times 10^{-5}\rm\,eV^2 \,,
\end{equation}

along with an unstable $\nu_2$ satisfying the lifetime constraint of
Eq.\ (\ref{RFlifetime}), explain all three neutrino anomalies
without violating any known data.  The very small entry, $\epsilon$
in Eq.\ (\ref{U}), is fixed by unitarity of the mixing matrix.

\medskip\noindent
\underline{\bf Big Bang Nucleosynthesis}

A neutrino with lifetime as short as the one discussed here decays
before the primordial neutrinos decouple (about 2 MeV). Then the
decay products would also achieve thermal equilibrium, increasing
the effective number of neutrino species $N_\nu$ to nearly 5, and
would be in conflict with the bound on light degrees of freedom
which is generally considered to be about 3.5 \cite{walker} or even
lower\cite{steigman}. Given all the uncertainties, more cautious and
conservative estimates\cite{kernan} lead to upper bounds on $N_\nu$
in the range of 4 to 5. Furthermore, the bound is modified if there
is lepton number asymmetry present\cite{foot}. Hence, although this
may be a potential problem, we do not regard it as fatal.

\medskip\noindent
\underline{\bf Supernova Emission}

The decay of $\nu_\mu$ at the fast rate envisaged here will
certainly modify supernova dynamics, since the decay will occur
inside the core.  In the Dirac case, the sterile decay products will
carry away energy on a very short time scale and this probably
conflicts with the observed period of a few seconds of the SN1987A
neutrino burst.  In the Majorana case, the decay products are not
sterile and this problem does not arise, and the effects on dynamics
should be much milder. It is not clear what is ``the precise range
of parameters that can be ruled out or ruled in by the SN1987A
signal''\cite{raffelt}.

\medskip\noindent
\underline{\bf Cosmic $\nu$ Fluxes}

In neutrinos coming from distant sources, such as Supernovae, AGN's,
and GRB's, the $\nu_\mu(\approx \nu_2)$'s have decayed away and only
$\nu_e, \bar\nu_e, \nu_\tau$ and $\bar\nu_\tau$ will arrive at the
Earth. Existing neutrino telescopes, as well as those under
construction, will not observe the tiny component ($\sim
\sin^4\theta\sim 10^{-3}$) of surviving $\nu_\mu$'s (apart from
atmospheric ones).

\medskip\noindent
\underline{\bf Future Tests}

There are several opportunities to test the neutrino decay
hypothesis decisively.  One is that the Super-Kamiokande
collaboration, with sufficient data, can distinguish between the two
$L/E$ distributions: the oscillatory one as given in
Eq.~(\ref{Pmumu-osc}) and the decaying one as given in
Eq.~(\ref{eq:Pmumu}).  A second is that in the upcoming $\mu$'s in
Super-K (which come from $\nu_\mu$'s of average energy around
100~GeV) the decay should have very little effect due to the long
lab--frame lifetime of the $\nu_2$ and the zenith angle distribution
of events should be undistorted.  For the same reason, partially
contained $\mu$'s should show some up-down asymmetry but less
pronounced than for the fully-contained events.

A third opportunity exists with future long--baseline
experiments\cite{long-b}.  Expectations are quite different with the
neutrino decay interpretation of Super-K results, compared to the
oscillation interpretation.  In the $\nu_\mu$ decay scenario the
typical $\nu_\mu$ survival probability is 94\%, the
$\nu_\mu$--$\nu_\tau$ conversion probability is about 6\%, and the
$\nu_\mu$--$\nu_e$ conversion probability about 0.2\%.(There is, in
the Majorana model, a further small $\nu_\tau$ flux from the decay
products; however their energies are much lower than the parent
$\nu_\mu$'s due to the backward-peaked nature of the decay.)

\medskip\noindent
\underline{\bf Conclusion}

We have presented a neutrino decay scenario capable of explaining
the atmospheric neutrino data. Furthermore, it enables us to explain
all neutrino anomalies with just three neutrino flavors.  The
scenario violates no available data, and is economical in that only a
single physics parameter ($\alpha =m_2/\tau_0$) is added beyond
those of the usual oscillation phenomenology.  The decay possibility
can be checked in future Super-K data as well as in forthcoming
long-baseline experiments and neutrino telescopes.

\medskip\noindent
\underline{\bf Acknowledgments}

We are grateful to Graciela Gelmini, ~Jim Pantaleone and
participants of the Trieste Relic Neutrino Workshop(September
16-19,1998) for clarifying and stimulating discussions. This work
was supported in part by the U.S.~Department of Energy and in part
by the Wisconsin Alumni Research Foundation.


\begin{references}
\frenchspacing
%\section*{References}
%\begin{thebibliography}{99}

\bibitem{super-k} Y. Fukuda et al., the Super-Kamiokande
Collaboration, Phys. Rev. Lett. {\bf 81}, 1562 (1998).

\bibitem{old} V. Barger and K. Whisnant, Phys. Lett. {\bf 209B}, 365
(1988); J.G.~Learned, S.  Pakvasa, and T.J.~Weiler, ibid. {\bf
207B}, 79 (1988); K.~Hidaka, M.~Honda, and S.~Midorikawa,
Phys. Rev. Lett. {\bf 61}, 1537 (1988).

\bibitem{ma} E. Ma and P. Roy, Phys. Rev. Lett. {\bf 80}, 4637
(1998); Y. Grossman and M. P. Worah, hep-ph/9807511; G. Brooijmans,
hep-ph/9808498; M. C. Gonzalez-Garcia et al., hep-ph/9809531.

\bibitem{britton} D. L. Britton et al., Phys. Rev. {\bf D49}, 28
(1994).

\bibitem{acker} A. Acker, S. Pakvasa, and J. Pantaleone,
Phys. Rev. {\bf D45}, R1 (1992).

\bibitem{bkp} V. Barger, W.-Y.~Keung, and S.~Pakvasa,
Phys. Rev. {\bf D25}, 907 (1982).

\bibitem{pdg98} Particle Data Group, C. Caso et al.,
Eur. Phys. J. {\bf C3}, 1 (1998).

\bibitem{bahcall} J.N.~Bahcall, P.~Krastev, and A.~Smirnov,
hep-ph/9807216; N.~Hata and P.~Langacker, Phys. Rev. {\bf D56}, 6107
(1997).

\bibitem{lsnd} C. Athanassopoulos et al. (LSND Collaboration),
Phys. Rev. Lett. {\bf 81}, 1774 (1998); ibid. {\bf 77}, 3082 (1996);
Phys. Rev. {\bf C54}, 2685 (1996).

\bibitem{gelmini} G. Gelmini and M. Roncadelli, Phys. Lett. {\bf
B99}, 411 (1981); Y. Chicashige, R. N. Mohapatra and R. Peccei,
Phys. Lett. {\bf B98}, 1980.

\bibitem{MajModel} A. Acker, A. Joshipura, and S. Pakvasa,
Phys. Lett. {\bf B285}, 371 (1992); G. Gelmini and J. W. F. Valle,
Phys. Lett. {\bf B142}, 181 (1984); K. Choi and A. Santamaria,
Phys. Lett. {\bf B267}, 504 (1991); A. S. Joshipura, PRL
Report-PRL-TH-91/6; A. S. Joshipura and S. Rindani, Phys. Rev. {\bf
D46}, 3000 (1992).

\bibitem{walker} T. P. Walker et al., Ap. J. {\bf 376}, 71 (1991).

\bibitem{steigman} G. Steigman, astro-ph/9803055.

\bibitem{kernan} S. Sarkar, hep-ph/9710273; P.J. Kernan and
S. Sarkar, Phys. Rev. {\bf D54}, R3681 (1996); C.J.~Copi,
D.N.~Schramm, and M.S.~Turner, Phys. Rev.  {\bf D55}, 3389 (1997);
B.D.~Fields et al., New Astron. {\bf 1}, 77 (1996); K.A.~Olive and
D.~Thomas, Astropart. Phys. {\bf 7}, 27 (1997).

\bibitem{foot} N. F. Bell, R. R. Volkas and Y. Y. W. Wong,
hep-ph/9809363 and references therein.

\bibitem{raffelt} G. G.  Raffelt, {\it Stars as Laboratories for
Fundamental Physics}, The University of Chicago Press, Chicago,
(1996) p. 564.
 
\bibitem{long-b} For World Wide Web links to these long-baseline and
other proposed neutrino oscillation experiments, see the Neutrino
Oscillation industry web page at
http://www.hep.anl.gov/NDK/Hypertext/nuindustry.html
\end{references}
\end{document}